\documentclass[twocolumn]{revtex4}
\usepackage{amsmath,amssymb,graphicx,bbm}

\newcommand{\mathi}{\mathrm{i}}
\newcommand{\tmop}[1]{\ensuremath{\operatorname{#1}}}
\newcommand{\um}{-}

\begin{document}

\title{Equilibrium thermodynamic properties of interacting two-component bosons in one dimension}

\author{Antoine Klauser${}^{1,2}$ and Jean-S\'ebastien Caux${}^2$}

\affiliation{${}^1$Instituut-Lorentz, Universiteit Leiden, P. O. Box 9506, 2300 RA Leiden, The Netherlands}
\affiliation{${}^2$Institute for Theoretical Physics, Universiteit van  
Amsterdam, 1018 XE Amsterdam, The Netherlands}
\date{\today}

\begin{abstract}
The interplay of quantum statistics, interactions and temperature is studied within the framework of
the bosonic two-component theory with repulsive delta-function interaction
in one dimension. We numerically solve the thermodynamic Bethe Ansatz and obtain the
equation of state as a function of temperature and of the interaction strength, the relative chemical potential
and  either the total chemical potential or a fixed number of particles, allowing to quantify the full crossover behaviour
of the system between its low-temperature ferromagnetic and high-temperature
unpolarized regime, and from the low coupling decoherent regime to the fermionization regime at high interaction.
\end{abstract}

\maketitle

\section{Introduction}

The increasingly common experimental realization of interacting quantum systems using cold atoms
has recently reignited interest in pushing our understanding of many-body physics beyond the traditional mean-field
level \cite{blochRoMP}.  This aspect is most in prominence in effectively one-dimensional realizations of bosonic
$^{87}\mbox{Rb}$ quantum gases with tunable interaction strength \cite{MoritzPRL91,LaburtheTolraPRL92,ParedesNATURE429,KinoshitaSCIENCE305,PolletPRL93},
for which the whole crossover from weakly- to strongly-interacting physics is accessible.

The case of locally interacting atoms confined to a uniform one-dimensional channel has the theoretical peculiarity  
of being integrable.  The simplest case of a single bosonic species defines the well-known Lieb-Liniger model
\cite{LiebPR130,LiebPR130b} for which a recent experimental study has shown that
the observed thermodynamic properties can be understood from the theory of integrable systems at finite
temperatures \cite{AmerongenPRL100}. In order to study even richer highly correlated systems,  multicomponent (spinor) gases have been experimentally realized \cite{LiaoNature,widera:140401,ErhardPRA}. 
This extension involves different hyperfine states which provide a pseudospin degree of freedom \cite{HarberPRA66,McGuirkPRL91, MertesPRL99, AndersonPRA80}. The control of the intra- and inter-species interaction strength via Feshbach resonances or 
state-dependent potentials \cite{Wicke1010.4545, HallerScience, ChinRMP82} opens the way for realizing a variety of integrable models.

The main interest in pursuing the study of multicomponent systems is that they provide situations where important interaction and quantum statistics effects coexist.  
These two aspects are not unrelated even in single-component systems:  as a simple
illustration, for a single bosonic species in one dimension, the limit of infinitely strong interactions (impenetrable bosons)
causes a crossover from bosonic to effectively fermionic behaviour \cite{TonksPR50,GirardeauJMP1}, at least
for physical quantities of density type.  Considering more than one component
however opens the door to much richer effects like the presence of spin wave excitations, with the possibility of crossover to many more regimes than the one component case. 

The study of multicomponent integrable systems really begins with the spin-$1/2$ fermion
problem \cite{McGuireJMP5,McGuireJMP6,McGuireJMP7,BrezinCRASPB263,GaudinPLA24}.  An interesting feature of
this system is that the attractive case has a correspondence to a bosonic system with twice the
interaction, in the sense that the equation for the ground state and particle energy coincide up to a sign \cite{GaudinPLA24}.  
A fundamental step forward was thereafter achieved by Yang, who showed that the repulsive delta-function
interaction problem admitted an exact solution irrespective of the symmetry requirement imposed on the
wavefunction \cite{YangPRL19}. For spin-$1/2$ particles, he showed that a generalized Bethe hypothesis
in the form of what is today called a nested Bethe ansatz could provide the system's wavefunctions, obtained the continuum equations
for the ground state, and calculated the general bound-state $S$-matrices \cite{YangPR168}.  
Sutherland \cite{SutherlandPRL20} generalized this to any irreducible representation of the permutation group,
so that in particular systems of type (in his notation) $B^x F^y$ with $x$ species of bosons and $y$ species of fermions
were amenable to an exact solution \cite{SutherlandBOOK}.  The ground state and excitations of 
multicomponent fermionic system were studied, both for repulsive and attractive interactions, by Schlottmann
\cite{SchlottmannJPC5,SchlottmannJPC6}.  This made extensive use of the `string hypothesis' for the
solutions to the Bethe equations, yielding the various dispersion branches in the repulsive case,
and the gapped color singlet ground states in the attractive one.

The bosonic multicomponent case has been less extensively studied.  For two components and in contrast with the Fermi gas,  the
ground state is (pseudospin) polarized as expected from basic arguments \cite{SutherlandBOOK} or more formally 
from a general theorem valid when spin-dependent forces are absent \cite{EisenbergPRL89}.  In pseudospin language, the ground
state is thus ferromagnetic, and the excitations at large coupling correspond to those in an
isotropic $XXX$ ferromagnetic chain \cite{GuanPRA76}, revealing thermodynamic properties which
are drastically different from those of the one-component Lieb-Liniger gas \cite{GuIJMPB16,LiEPL61}. 
Furthermore, recent studies of the ferromagnetic ground state revealed different dispersions for the charge and pseudospin excitations which therefore exhibits a spin-charge separation \cite{KleineNJP,FuchsPRL}.
The regime of strong interaction is still not completely understood and even if Girardeau's Fermi-Bose mapping has been showed and used for the study of the 1D spinor Bose gases
 \cite{DeuretzbacherPRL100} leading to a paramagnetic Tonk-Girardeau regime \cite{GuanPRA76}, this approach fails to provide first correction terms to this fermionization regime since the discernability of the bosons is missing.

The purpose of our paper is to further stimulate the contact between integrability theory and
experiments on multicomponent cold atoms, by providing quantitative predictions for the equation
of state and population densities of two-component interacting Bose gases as a function of temperature, interaction strength
and of either the available chemical potentials or the chemical potential difference and a fixed density of particles.
This work broadens and extends our earlier paper \cite{CauxPRA}.
 The paper is organized as follows.  In section \ref{sec:SETUP},
after defining our notations, we quickly review the construction of the eigenstates of the theory using
the Bethe Ansatz, and how these can be used to obtain the thermodynamics of the system in the
continuum limit via the solution of a (infinite) set of coupled integral equations.  Section \ref{sec:NUM}
outlines the method we have used to solve this system numerically, using two different approaches
allowing cross-checking of the results.  Section \ref{sec:TEMP} discusses the effect of the thermal fluctuation over the ferromagnetic ground-state, whereas section \ref{sec:INTERM} provides results
on the more challenging intermediate regimes.  Section \ref{sec:DECOH} discusses the results in the decoherents regime of low coupling where we compare the numerical results with a perturbative result and section \ref{sec:TONKS} presents the results at strong coupling where the gas enters the fermionization regime. We end with conclusions and
perspectives.  

\section{Setup}
\label{sec:SETUP}
Consider a collection of bosonic atoms of equal mass but having an internal $SU(2)$ degree of freedom (in practice, this would be for example two distinguishable hyperfine states, and could be thought
of as a (pseudo-)spin$-1/2$. The unique feature differentiating from a single species is the fact that distinguishability imposes symmetry of the many-particle wavefunction on the same-spin particles only.  For definiteness, we consider a one-dimensional ring of length $L$, in which a total
of $N$ atoms circulate.  The first-quantized Hamiltonian of the
system includes a free dynamical term to which a spin-blind interaction term is added, and reads
\begin{equation}
  \mathcal{} \mathcal{H}_N =^{} - \frac{\hbar^2}{2m} \sum^N_{i = 1} \frac{\partial^2}{\partial
  x^2_{i^{}}} + g_{1D} \sum_{1 \leq i < j \leq  N} \delta (x_i - x_j).
\label{eq:H_N}
\end{equation}
The effective one-dimensional coupling parameter $g_{1D}$ is related to the effective 1D scattering length $a_{1D}$
\cite{OlshaniiPRL81} via the relation $g_{1D} = \hbar^2 a_{1D}/2m$.  Hereafter, we will use the effective interaction
parameter $c = g_{1 D} m/\hbar^2$, and adopt the traditional convention of setting $\hbar = 2m = 1$ to simplify the notations.  
Note that this choice of interaction term involves fine-tuning two parameters:  more generally, we could have different intra- and inter-species scattering
lengths.  To preserve integrability however, these must all be equal.

Specializing to $N$ atoms of which $M$ have (in the adopted cataloguing) spin down, the Bethe Ansatz
provides eigenfunctions fully characterized by sets of rapidities (quasi-momenta) $k_j$, $j = 1, ..., N$ and pseudospin rapidities
$\lambda_{\alpha}$, $\alpha = 1, ..., M$, provided these obey the $N + M$ coupled equations \cite{YangPRL19,SutherlandPRL20}
\begin{eqnarray}
&&  e^{\tmop{ik}_j L}  =  - \prod_{l = 1}^N \frac{k_j - k_l + \tmop{ic}}{k_j -
  k_l - \tmop{ic}} \prod^M_{\alpha = 1} \frac{k_j - \lambda_a -
  \frac{\tmop{ic}}{2}}{k_j \um \lambda_a + \frac{\tmop{ic}}{2}}, \nonumber \\
%&& \hspace{5cm} j = 1,  \ldots, N \nonumber\\
&&  \prod^N_{l = 1} \frac{\lambda_{\alpha} - k_l -
  \frac{\tmop{ic}}{2}}{\lambda_{\alpha} - k_l + \frac{\tmop{ic}}{2}} = -
  \prod^M_{\beta = 1} \frac{\lambda_{\alpha} - \lambda_{\beta} -
  \tmop{ic}}{\lambda_{\alpha} - \lambda_{\beta} + \tmop{ic}}, 
%  \nonumber \\
%&& \hspace{5cm} \alpha = 1, \ldots, M 
\label{eq:Bethe_eq}
\end{eqnarray}
for $j = 1, ..., N$ and $\alpha = 1, ..., M$.  For a generic eigenstate, the solution to the Bethe equations
is rather involved and cannot be obtained in closed form.  Two observations allow to push the treatment further:  first, for $c>0$, the 
$k_j$ rapidities live on the real axis.  This is not true of the $\lambda_{\alpha}$ which are found to be
generically complex, but arranged into regular patterns called strings \cite{BetheZPA,TakahashiPTP}.  An $n$-string of $\lambda$'s 
is a congregation of $n$ rapidities sharing the same real value and having an even spacing of height $c$ in the imaginary direction.
The adopted notation for the $a$-th member of a $n$-string labeled by the index
$\alpha$ and centered on $\Lambda^n_{\alpha}$ is thus $\lambda^{n,a}_{\alpha} = \Lambda^n_{\alpha} + i \frac{c}{2}(n + 1 - 2a)$,
with the equality being exact only up to deviations which (according to the traditional string hypothesis) vanish in the 
infinite size limit.  Throughout our work, we will adopt this as a working hypothesis.  
Since the total number of each type of string is conserved under time evolution, each string type represents a quasiparticle of the theory.  
In the thermodynamic limit, the distribution of all rapidities can be encoded into a set of smooth functions representing the densities of roots for each string type.
The Bethe equations then become a set of coupled integral equations for (quasi)particle and (quasi)hole root distribution functions.
We refer the reader who is unfamiliar with these to our summary of important formulas in appendix ( \ref{sec:TBA} ).

The Thermodynamic Bethe Ansatz (TBA) allows to exploit the condition of thermal equilibrium \cite{TakahashiBOOK,yang&yang} to 
obtain the Yang-Yang-Takahashi (YYT) like equations \cite{yang&yang,GuanPRA76,GuIJMPB16} for $\epsilon (\lambda)$, the dressed energy, and $\epsilon_n ( k)$, length-$n$ string dressed energy, $n = 1, 2, ...$
\begin{eqnarray}
  \varepsilon ( k) \!\!&=&\!\!  k^2 - \mu - \Omega - T a_2 \!\ast\! \ln
  \left[ 1 + e^{-\varepsilon( k)/T} \right] \nonumber \\
  && - T \sum_{n = 1}^{\infty} a_n \!\ast\! \ln \left[ 1 + e^{- \varepsilon_n( k)/T} \right] \nonumber\\
  \frac{\varepsilon_1 ( k)}{T} \!\!&=&\!\! f \!\ast\! \ln \left[ 1 + e^{-\varepsilon( k)/T} \right] + f \!\ast\! \ln \left[ 1 + e^{\varepsilon_{2}( k)/T} \right], \nonumber\\
  \frac{\varepsilon_n ( k)}{T} \!\!&=&\!\! f \!\ast\! \ln \left[ 1 + e^{\varepsilon_{n + 1}( k)/T} \right] + f \!\ast\! \ln \left[ 1 +
  e^{\varepsilon_{n - 1}( k)/T} \right]
  \nonumber \\ &&\hspace{5cm} (n > 1),
\label{eq:system}
\end{eqnarray}
with the standard convolution notation $g \ast h ( k) \equiv \int_{-\infty}^{\infty} d k' g ( k -  k')
h ( k')$, and the kernels $a_n ( k) = \frac{1}{\pi} \frac{n c / 2}{(n c / 2)^2 +  k^2}$ and
$f ( k)  = \frac{1/2c}{\cosh ( \pi  k/c)}$.  The set of coupled equations is completed
with the asymptotic conditions
\begin{equation}
  \lim_{n \rightarrow \infty} \frac{\varepsilon_n ( k)}{n} = 2 \Omega
\label{eq:epsnlarge}
\end{equation}
for high-level functions. From these can be derived the large-rapidity asymptotes 
\begin{equation}
  \lim_{ k \rightarrow \infty} \varepsilon_n ( k) \equiv
  \varepsilon_n^{\infty}, \hspace{1cm} n = 1, 2, ...
\label{eq:epslambdalarge}
\end{equation}
for the large rapidity asymptotic values of the individual functions, where we have defined the numbers 
\begin{equation}
\varepsilon_n^{\infty} \equiv 2 \Omega n + T \ln \left( \left( \frac{1
  - e^{- \frac{2 \Omega}{T} (n + 1)}}{1 - e^{- \frac{2 \Omega}{T}}}
  \right)^2 - e^{- \frac{2 \Omega}{T} n} \right)\hspace{0.5cm}. \label{eq:epsninfty}
\end{equation}

The thermodynamics of the system is provided by the solution set of dressed energies as a function of the temperature $T$, the
total $\mu$($=\frac{\mu_1+\mu_2}{2}$ with $\mu_i$ the chemical potential specific to the $i$th component) and relative $\Omega$ ($=\frac{\mu_1-\mu_2}{2}$) chemical potential (see appendix \ref{covariance} concerning the $c$ parameter). The Gibbs free energy per unit length is given by
\begin{equation}
 g =  - \frac{T}{2 \pi} \int^\infty_{-\infty} \ln \left[ 1 + e^ {-
  \varepsilon ( k)/T} \right] d k
\label{eq:g}
\end{equation}
while the linear density of the $i$th boson component is
\begin{equation}
  n_i  =  - \frac{1}{2} \left( \frac{\partial g}{\partial \mu} + (-1)^{i-1}
  \frac{\partial g}{\partial \Omega} \right)\hspace{0.5cm}.
\label{eq:n_i}
\end{equation}
The entropy density is given by the standard thermodynamic identity 
\begin{equation}
 s  =  -\frac{\partial g}{\partial T}
\label{eq:s}
\end{equation}
and the local density-density correlator \cite{KheruntsyanPRL,GangardtPRL} is given (using the Hellman-Feynman theorem) by
\begin{eqnarray}
g^{(2)}& = \frac{\sum_{i,j}\langle \Psi^\dagger_i\Psi^\dagger_j \Psi_j\Psi_i \rangle}{( \sum_{i}\langle \Psi^\dagger_i \Psi_i\rangle )^2}& =  \frac{ \frac{-\partial g}{\partial c}}{ ( \sum_{i}n_i ) ^2} \hspace{0.5cm}.
\label{eq:g^2}
\end{eqnarray}
\section{Numerical treatment}
\label{sec:NUM}
To solve the infinite system of transcendental coupled equations (\ref{eq:system}), we have developed two different numerical algorithms. We can then independently check the results by comparison.
% and we access a broader range of parameters (their efficiency should be different with the parameters region).
We first discuss the common approach for the numerical treatment. Afterwards, we will describe
and motivate the choices we made to build the two algorithms. \

Two cutoffs are applied on the system (\ref{eq:system}) to implement a numerical process. Firstly we reduce the number of
functions for computing to $n_{\max}$, replacing for $n > n_{\max}$ these functions by their asymptotic value (limit $n \rightarrow \infty$ in (\ref{eq:epsnlarge})). Secondly, we reduce the range of integration of the convolution to a finite value. Supposing that as
$k \rightarrow \pm \infty$, $\varepsilon (k) \sim k^2$ and that $\varepsilon_n$
becomes $\varepsilon^\infty_n$ (\ref{eq:epslambdalarge}), we limit the range of
the $k$ values \ to $[- \Delta_n : \Delta_n]$. In consequence, we estimate the
solution by 1+$n_{\max}$ functions $\{\varepsilon, \varepsilon_1,
\varepsilon_2, \ldots, \varepsilon_{n_{\max}} \}$ and we evaluate them by $N_i$ points over the interval $[- \Delta_i :
\Delta_i]$ ($i = 0, \ldots, n_{\max} $). To compute the solution we proceed
by iterations, starting from the free bosons form of $\varepsilon$ and the
asymptotic values $\varepsilon_n^{\infty}$.

The previous paragraph describes how to get the particle dressed energy, $\varepsilon( k )$, 
and consequently allows one to compute the Gibbs free energy (\ref{eq:g}). But in order to
compute any other thermodynamic quantity involving a derivative of $g$ (\ref{eq:n_i}, \ref{eq:s}, \ref{eq:g^2}), we use the system for the corresponding set of functions $\{
\frac{\partial \varepsilon}{\partial var}, \frac{\partial
\varepsilon_1}{\partial var}, \ldots\}$ with $var \in \{\mu, \Omega, T, c \}$ (\ref{eq:dnu_system},  \ref{eq:dT_system}, \ref{eq:dc_system}).
We achieve the numerical solutions by the same method of discretisation introduced before and
using the set $\{\varepsilon, \varepsilon_1, \ldots, \varepsilon_{n_{\max}}
\}$ solution of (\ref{eq:system}). It would be possible to
numerically differentiate $G$ or $\varepsilon$ to compute these quantities.
This method would however achieve only much less accuracy for given computational effort.

\subsubsection{Fixed density of particles}

For physical interpretation of the results and for the identification of the different regime crossovers,
results with fixed density of particles could be more convenient. For this purpose, 
we implemented a Newton's method on top of our main algorithms that finds the chemical potential, $\mu$, corresponding to a desired density, $n_1+n_2$.

\subsubsection{Accuracy and precision}

By the use of the method mentioned above to solve the system, we are
confronted with two limitations on exactitude. The first comes form the numerical approximation of
the system: discretization of the functions, limitation of the integration range and the
number of function leads to an imprecision on the results.
Second, the fact that we solve the system by iteration approaching but not reaching the solution, leads to a inaccuracy.% (note that we make a difference between inaccuracy and imprecision). 

The imprecision is limited as described hereafter. The discretisation of
the functions adds an error $\mathcal{O}( \frac{1}{N^2})$ in the convolutions
($N$ being the numbers of points) which induces an imprecision that one can easily keep low.
Concerning the range cutoff, all the thermodynamic quantities are
$\propto e^{- \varepsilon (k) / T}$, with $\varepsilon (k) \sim k^2$ when $k
\gg 1$. Therefore we reduce this effect on the results by taking an appropriate $k$ spacing such that $\Delta ^2 /T\gg 1$. Finally, as we can see
from (\ref{eq:system}, \ref{eq:dnu_system},\ref{eq:dT_system},\ref{eq:dc_system}), all the contributions
of the $\varepsilon_n$ functions in $\varepsilon (k)$ are $\propto e^{-
\varepsilon_n (k) / T}$. Knowing that for $n \gg 1$, $\varepsilon_n (k)
\cong 2 n \Omega$, we therefore keep this imprecision small by choosing $\frac{n_{\text{max}}\Omega}{T} \gg 1  $. In the results
shown in all plots, the precision of the results is estimated to always be much smaller than the width of
the curves.

The problem is different for the accuracy. Supposing that the
solutions $\{\varepsilon, \varepsilon_1, \varepsilon_2, \ldots\}$, $\{
\frac{\partial \varepsilon}{\partial var}, \frac{\partial
\varepsilon_1}{\partial var}, \ldots\}$ ($var \in \{\mu, \Omega, T, c \}$) exist, we suppose that by iteration we approach these solutions but the distance from the solution is nevertheless 
unknown. We estimate the accuracy empirically. Increasing the
total number of iterations exponentially, we observe the convergence of the
results and judge the accuracy value. In the following results, the accuracy
is estimated to be of order of the line width and therefore globally is the higher limitation on exactitude of the results.

\subsection{FFT based algorithm}

\subsubsection{Idea}

During the iterative process, the major part of calculation time is taken by the
evaluation of each convolution. Indeed by calculating the integrals by simple
trapezoidal sums, this charge represents $\sim \mathcal{O}(N_iN_{i\pm 1})$ operations for each function. Starting from this observation, the basic idea
of this algorithm is to use the Fast Fourier Transform to calculate the
convolution: $(f \ast g) (x) = \tmop{FT}^{- 1} (\tmop{FT} (f) \cdot \tmop{FT}
(g))$ (with $\tmop{FT}$: Fourier Transform). The computing time of each
convolution using FFT is thus only $\sim \mathcal{O}(N_i^{} \log (N_i))$. The conditions of use of the FFT are that the functions $f$ and $g$ must be integrable
and that the values of these functions must be zero outside $[- \Delta_i : \Delta_i]$. This could be easily achieved by treating the constant part of the function separately from the nontrivial part.
Moreover this method imposes the numerical constraints that all the points must be equally spaced and that the range and the numbers of points of each function must be the same.

\subsubsection{Practically}

As a consequence of this, we have a set of $n_{\max} + 1$ functions each to be evaluated on
$N$ points and within the range $[- \Delta, \Delta]$. We start by setting up the system with three arbitrary parameters: $(D_0, \Delta_0,
n^0_{\max})$ with $D = \frac{N}{2 \Delta}$. During the convergence, we adjust them dynamically with the use of
the following precision indicators. We firstly estimate the $n$th iteration precision with
\begin{widetext}
\begin{equation}
  \sigma_{\tmop{it}} = - \frac{T}{2 \pi G} \frac{2\Delta}{N}\sum^N_{i = 1}  \left | \ln
  \left[ 1 + \exp (- \frac{\varepsilon^{(n)} (k_i)}{T}) \right] - \ln \left[ 1 +
  \exp (- \frac{\varepsilon^{(n-1)} (k_i)}{T}) \right] \right |
\end{equation}
with $\varepsilon^{(n)} (k_i)$ being the value of $\varepsilon (k_i)^{}$ after $n$
iterations. This formula has to be understood as the variation of the Gibbs free energy between two steps (see
(\ref{eq:g})). We measure similary how the
parameters $n_{max}$ and $\Delta$ influence the precision with these two indicators
\begin{eqnarray}
  \sigma_{n_{\max}} & = & - \frac{T}{2 \pi G} \frac{2\Delta}{N} \sum^N_{i = 1} \left|
  \ln \left[ 1 + \exp (- \frac{\varepsilon_{n_{\max}} (k_i)^{}}{T}) \right] -
  \ln \left[ 1 + \exp (- \frac{\varepsilon_{n \rightarrow \infty}^{}}{T})
  \right] \right | \\
  \sigma_{\Delta} & = & - \frac{T 2 \Delta}{2 \pi n_{\max} G} \sum^{}_n \left |
  \frac{1}{2} \ln \left[ \left(1 + \exp (- \frac{\varepsilon_n (k_1)}{T})
  \right) \left( 1 + \exp (- \frac{\varepsilon_n (k_N)^{}}{T}) \right) \right]
   - \ln \left[ 1 + \exp (- \frac{\varepsilon_n^{\infty}}{T}) \right]
  \right |
\end{eqnarray}
\end{widetext}
During computation, if $\sigma_{n_{\max}} >
\sigma_{\tmop{it}}$, $n_{\max}$ is increased and  if $\sigma_D >
\sigma_{\tmop{it}}$,$\Delta$ is lengthened. The precision related by the density of points, $D$, is hard to quantify but can be estimated by cross-checking with the second method which has a non-uniform distribution of points. 
We then increase it step by step as one goes along the iterating process.

Once  satisfactory values for $\sigma_{\tmop{it}}^{\tmop{sol}}, D^{\tmop{sol}}$ are achieved, we assume the solution has been reached and we calculate the Gibbs free energy from (\ref{eq:g}). The derivatives of the Gibbs free energy are computed using the derivative systems
(\ref{eq:dnu_system},  \ref{eq:dT_system},  \ref{eq:dc_system}) with the final $(\Delta, n_{\max}, D^{\tmop{sol}})$
determined by the first iterative process and we approximate the precision
during the iterations by
\begin{eqnarray}
  \sigma_{\tmop{it}}^{var} & = & \frac{1}{2 \pi \frac{\partial G}{\partial var}}
  \sum^N_{i = 1} \Delta_k \frac{| \frac{\partial \varepsilon (k_i)^n}{\partial
  var} - \frac{\partial \varepsilon (k_i)^{n - 1}}{\partial var} |}{1 + \exp
  ( \frac{\varepsilon (k)^n}{T})}, \nonumber\\
&& var \in \{ \mu, \Omega, T, c \}.
\end{eqnarray}
Since the arbitrary $\sigma_{\tmop{it}}^{var, \tmop{sol}}$ with $D^{\tmop{sol}}$ are
reached, we consider that we have a good evaluation of the solutions.

\subsection{Flexible-density method}
\label{subsec:FLEX}
A second, completely independent implementation of the numerical solution to the
coupled integral equations has been pursued as part of our work.  Here, we do not
make use of the fast Fourier transform, but rather maintain total flexibility in
the choice 1) of density of sampling points within each function, 2) of the $\Delta_i$
limits used at each level, 3) of the relative total number of points used at each
level, and 4) of the total number of functions used.  This advantage allows one to 
concentrate computational resources where they are needed, but comes at the cost
of being only able to perform convolutions between {\it e.g.} levels $i$ and $j$ at speed of 
order $N_i N_j$ where $N_i$ is the number of points used at level $i$.  This second
algorithm performs more or less equally well as the first, and allows one to certify the results obtained.  

In summary, this second algorithm works as follows.  Depending on the physical
parameters requested, an initial choice is made of the number $n_{\max}$ of functions
to be considered, and of the limits $\Delta_i$ at each level.  A dynamical parameter
called the running precision is initialized, which estimates the numerical accuracy obtained
in computing the free energy using the points configuration used.  The coupled
equations are then iterated (possibly using extrapolations) in order to achieve a
certain degree of convergence, measured by the condition that the total rate of flow of all points as the
iterations proceed becomes smaller than the running precision.  

At this point, a {\it cycle} is initiated.  This entails a number of steps, with the
objective of increasing the accuracy, {\it i.e.} of decreasing the running precision
achieved.  First, each function is examined in turn, and points are added in regions
with larger curvature.  Second, the limits $\Delta_i$ are extended (and points added)
if the value of the function at the previous limit is not sufficiently close to its
analytically-determined asymptotic value $\epsilon^{\infty}_i$.  Third,
new functions are added ({\it i.e.} $n_{\max}$ is increased) if the highest function is not
sufficiently close to its asymptotic value throughout the $k$ line. A new value of the running precision is then
determined, based on the refinements just performed on the distribution of points.
Finally, iterations are performed until the flow rate drops below this running precision.

For a specific set of physical parameters, a total allowed time is also given to the
program.  This second implementation then performs cycles one after the other, yielding
increasingly accurate results, until the allowed time is exhausted.  An estimate of the
absolute accuracy of the whole procedure can thus be obtained by comparing the results 
from runs with different total allowed times.
\section{Quantum statistics versus temperature fluctuations}
\label{sec:TEMP}
The $SU(2)$ degree of freedom in combination with the
bosonic statistics in 1D leads to a macroscopic behavior: the polarization of the ground state.
This phenomenon which occurs in every bosonic system with no explicit component-dependent forces, has been already proven in the literature \cite{EisenbergPRL89}. We will give here an interpretation in terms of the string structure of the Bethe solutions and provide a quantitative result for the persistence of this effect for non-zero temperature.

The polarization at zero temperature can be directly linked to the underlying string structure of the Bethe equation solutions.
In eq. (\ref{eq:system}), the $\varepsilon(k)$ function depicts the charge degree of freedom and the $\varepsilon_n (k)$ functions  the spin degrees of freedom. Moreover those latter functions express the dynamics of quasiparticle forming a colour 2 energetically disfavored state made of $n$ particles. %describe bound state in one colour component that are present in the wave function and 
As the temperature of the system goes down, the contribution of these states in the equilibrium decreases. 

\begin{center}
\begin{figure}[h]
  \includegraphics[width=0.5\textwidth]{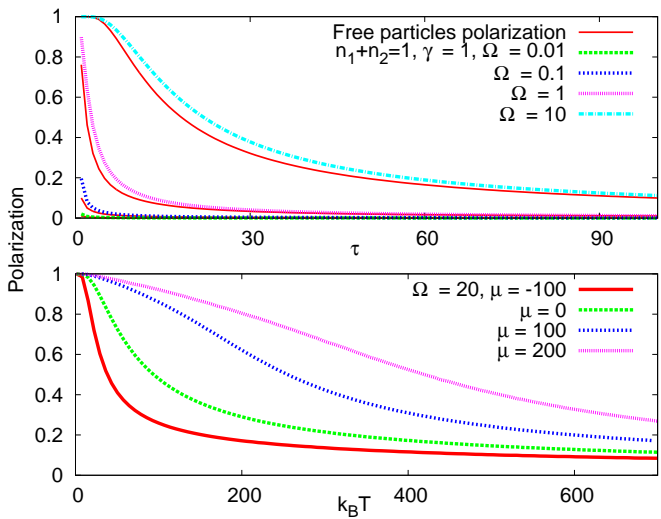}
  \caption{\label{fig:T} The top graph shows polarization of the 2CBG , $(n_1-n_2)/(n_1+n_2)$, for fixed density of particles and interaction strength
 ($\gamma$) as a function of the reduced temperature ($\tau$). The set of curves for several different $\Omega$ are compared to the curves of a free gas. 
 The bottom plot shows the isobar polarization as a function of the temperature for four different chemical potentials: \{$-100,0,100,200$\}.
In this latter graph, the density $\gamma$ and $\tau$ vary along the curves.}
\end{figure}
\end{center}

We hereafter explain how at the limit $T = 0$ there only
remains one spin-gapped state gathering all particles in the 1st component.
In the YYT equations (\ref{eq:system}) where the $\varepsilon_n (k)$ %functions represent 
are the dressed energies of an $n$-string% of component $2$. From this point of view
, a phenomenological approach to $T = 0$ is possible. Taking the first line of
(\ref{eq:system}) and approximating the values $\varepsilon_n (k) \approx 2 n\Omega$ 
, if one takes the limit $T \rightarrow 0$, $ T \sum_{n =
1}^{\infty} \ln \left[ 1 + \exp (- 2 \frac{n \Omega}{T}) \right ] \rightarrow 0$ as $\Omega$ is defined positive.
The contribution of the colour 2 particles then disappears from the thermalized state and this reveals that the colour 1 component drives away all colour 2 particles,
 forming a fully polarized spin-gapped state. (In the Bethe equations, only the 2nd component part of the wave function is represented by quasiparticles).
We show this expelling in figure \ref{fig:T} where we plot polarization curves (bottom set) as a function of $\mu$.
 As the chemical potential increases, the interaction parameter, $\gamma$ (=$\frac{c}{n_1+n_2}$) decreases monotically and the polarization persists to higher temperatures.
In \cite{FuchsPRL}, Fuchs et al. revealed that the effective mass of an isospin wave above the polarized ground state is very high in the strong coupling regime.
Furthermore, it is surprising to see that when $\mu$ increases, the polarized ground state is more resistant to thermal fluctuation even though the isospin wave mass decreases.

In the context of spontaneous imbalance in binary mixtures \cite{CazalillaPRL,KolezhukPRA,TakayoshiPRA}, it has been shown that at zero temperature, a mixed gas is unstable and exhibits a spatial phase separation.
 But no quantitative predictions have been made for finite temperature in 1D. From figure \ref{fig:T}, we see that the polarization remains at higher temperatures for high value of $\mu$. By qualitative identification of the ferromagnetic behavior with the the spatial demixing, we can speculate that a phase separation would resist better to temperature in the low coupling regime than for $\gamma\gg 1$.

\begin{center}
  \begin{figure}[h]
    \includegraphics[width=0.5\textwidth]{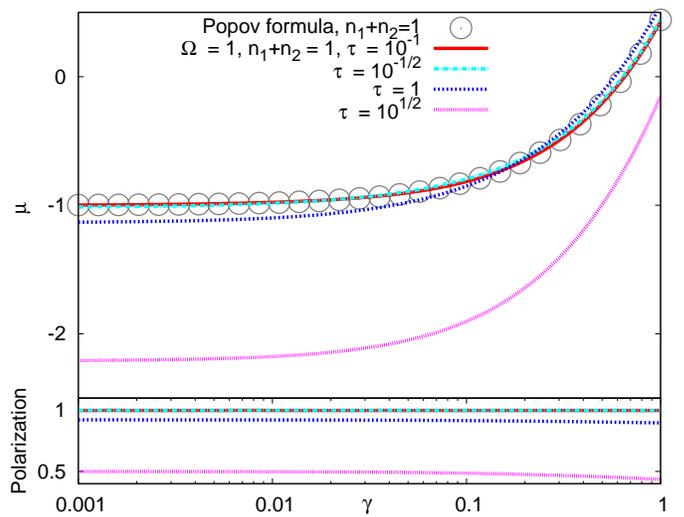}
    \caption{\label{fig:Popov}For $\gamma \ll 1$, and $\tau \ll 1$, the 2CBG chemical potential follows the Popov's expression for the zero temperature Lieb-Liniger case.}
  \end{figure}
\end{center}

The figure \ref{fig:T} shows as well curves of polarization for fixed particle density, interaction strength and fixed $\Omega$ as a function of the reduced temperature, 
$\tau= \frac{T}{T_D}$ where $T_D= (n_1+n_2)^2$ is the degeneracy temperature \cite{KheruntsyanPRL}.
We compare these results with the polarization of an interaction-free 2CBG which is $\frac{n^0_1-n^0_2}{n^0_1+n^0_2} = \tanh(\Omega / k_B T)$.
The non-negligeable difference which appears between the curves of same $\Omega$ is the ferromagntic effect which is a consequence of the bosons interaction. 
As $T$ approaches zero, the gas polarizes and the remaining component behaves like a Lieb-Liniger gas of chemical potential $\mu_1 = \mu + \Omega$.
The results are then comparable to the results of V. N. Popov \cite{PopovTMP} for the density of the
Lieb-Liniger gas at $T = 0$ and $\gamma \ll 1$:
\begin{equation}
  \rho (\mu) = \frac{\mu}{2 c} + \frac{\sqrt{\mu}}{\sqrt{2} \pi} + c (
  \frac{1}{2 \pi^2} - \frac{1}{24}) + \ldots
\end{equation}
As shown in Fig. \ref{fig:Popov}, by lowering down the reduced temperature of the gas to $\tau \ll 1$, the total chemical potential
corresponds to the zero temperature low coupling regime formula. Moreover, in this particular regime, the corresponding polarizations lines turn out to be almost constant along $\gamma$. 

\begin{center}
\begin{figure}[h]
  \includegraphics[width=0.5\textwidth]{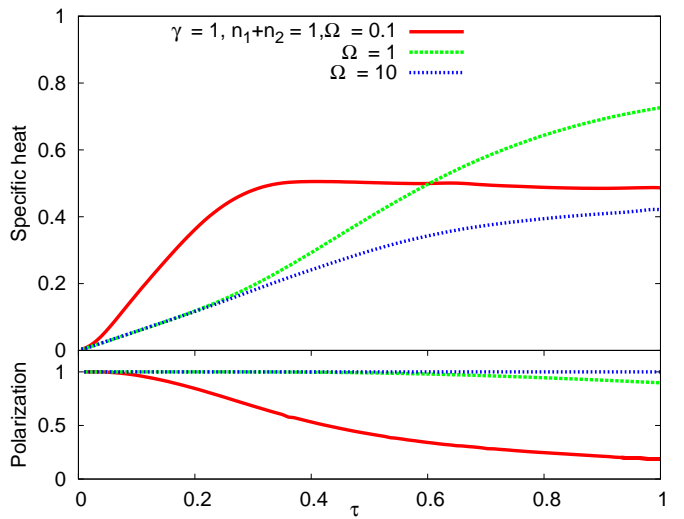}
  \caption{\label{fig:C_lowT} The specific heat of the 2CBG for fixed density of particles and interaction strength as a function of the reduced temperature.
The different chemical potentials of each curve have a value closed to $\gamma$. Similarly to a free 2CBG case, we see a peak in the specific heat whos the position depends on $\Omega$.}
\end{figure}
\end{center}

\begin{center}
\begin{figure}[h]
  \includegraphics[width=0.5\textwidth]{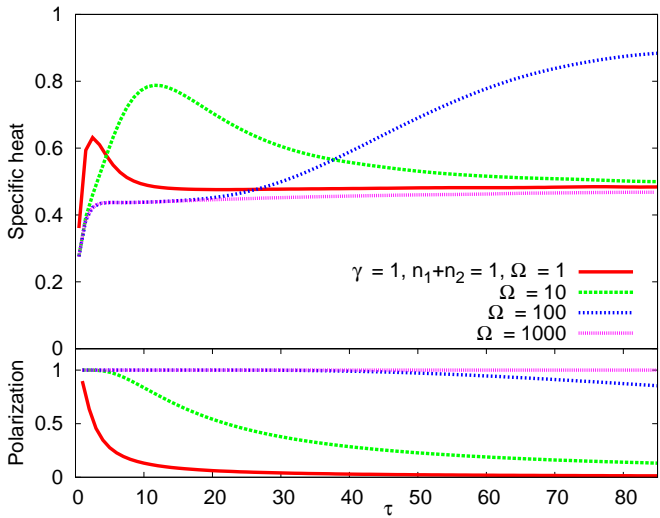}
  \caption{\label{fig:C} The specific heat of the 2CBG for fixed density of particles and interaction strength as a function of the reduced temperature.
The different chemical potentials have a value $\Omega \gg \gamma$. In the limit of high temperature, the gas becomes ideal and the specific heat takes the value $1/2$.}
\end{figure}
\end{center}

The specific heat capacity of the gas at fixed density of particles which is accessible via the entropy (\ref{eq:s})
provides a view on the thermal degrees of freedom of the system.
Figure \ref{fig:C_lowT} shows that in a 2CBG at low temperature with strong $\Omega$, the heat capacity is similar to that of a Lieb-Liniger Bose gas contributed by phonons. 
If the relative chemical potential is lower or of order of the temperature, the two component degree of freedom appears and creates peaks similarly to a paramagnetic spinor Bose gas \cite{GuanPRA76}. The maximum in the specific heat moves in higher temperature as the relative chemical potential increases.
The higher temperature results are shown in \ref{fig:C} where we can observe that the peaks are located when $\Omega \sim T$. At high temperature the gase becomes decoherent classical
 (see \ref{sec:DECOH}) and the specific heat converges to $1/2$ which is the value of the simple 1D ideal gas.

Experiments trapping Helium$-4$ fluid into 1D nanopores \cite{WadaPRL} provide a possible realization of the 1D Bose gas and give access to measurement of the heat capacity of the unidimensional system. A similar realization with an isospin-$1/2$ might be possible and provide measurement of the 2CBG heat capacity.

\section{Results in the intermediate regime}
\label{sec:INTERM}
\begin{figure*}[]
  \begin{center}
    \includegraphics[width=\textwidth]{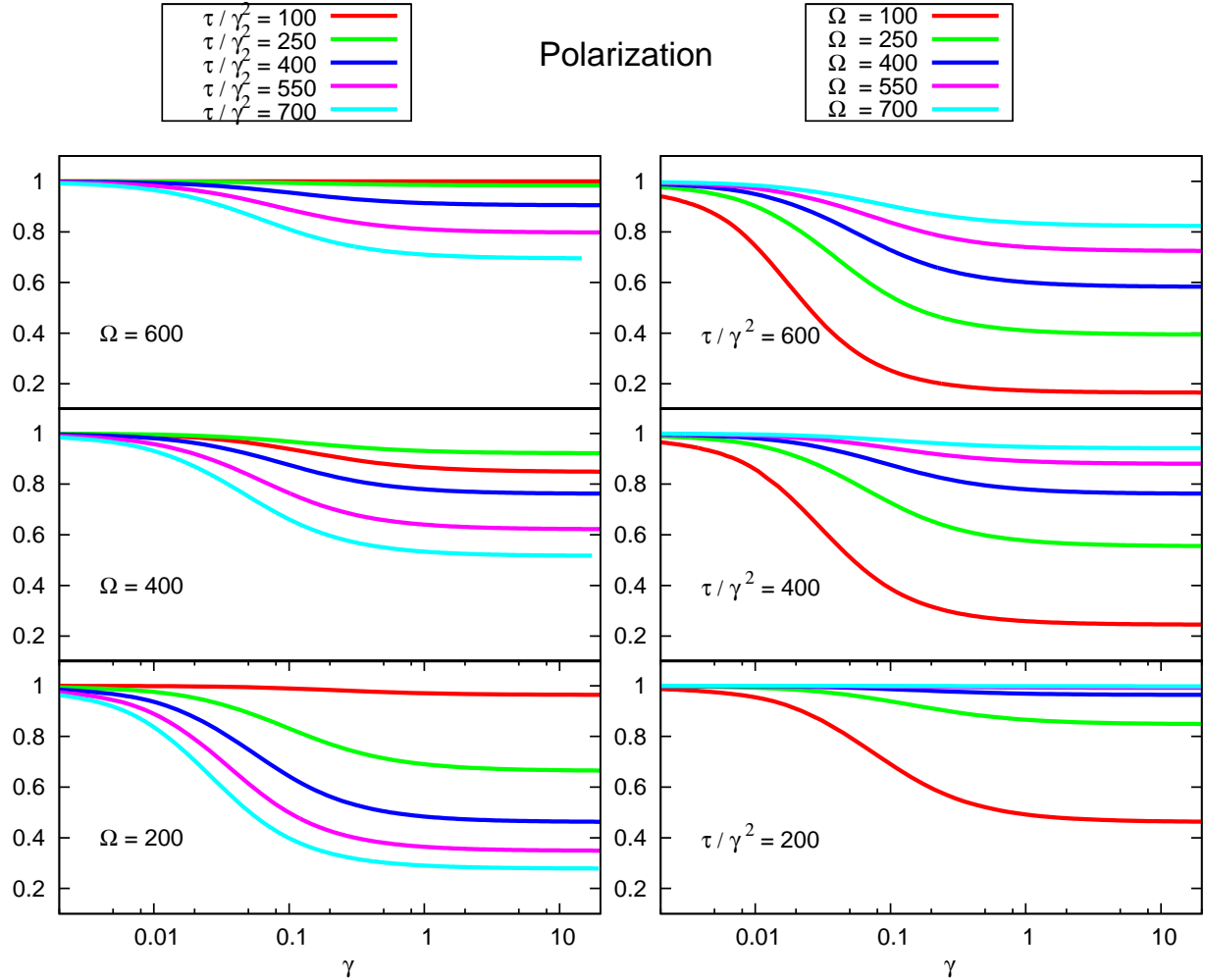}
       
  \caption{\label{fig:table_Pol}Polarization of the 2 component Bose gas
      as a function of the interaction strength $\gamma$  for fixed values of
      $\Omega (= \frac{\mu_1 - \mu_2}{2})$ and for fixed ratios $\tau / \gamma^2$.}
\end{center}
\end{figure*}
\begin{figure}[h]
  \begin{center}
    \includegraphics[width=0.5\textwidth]{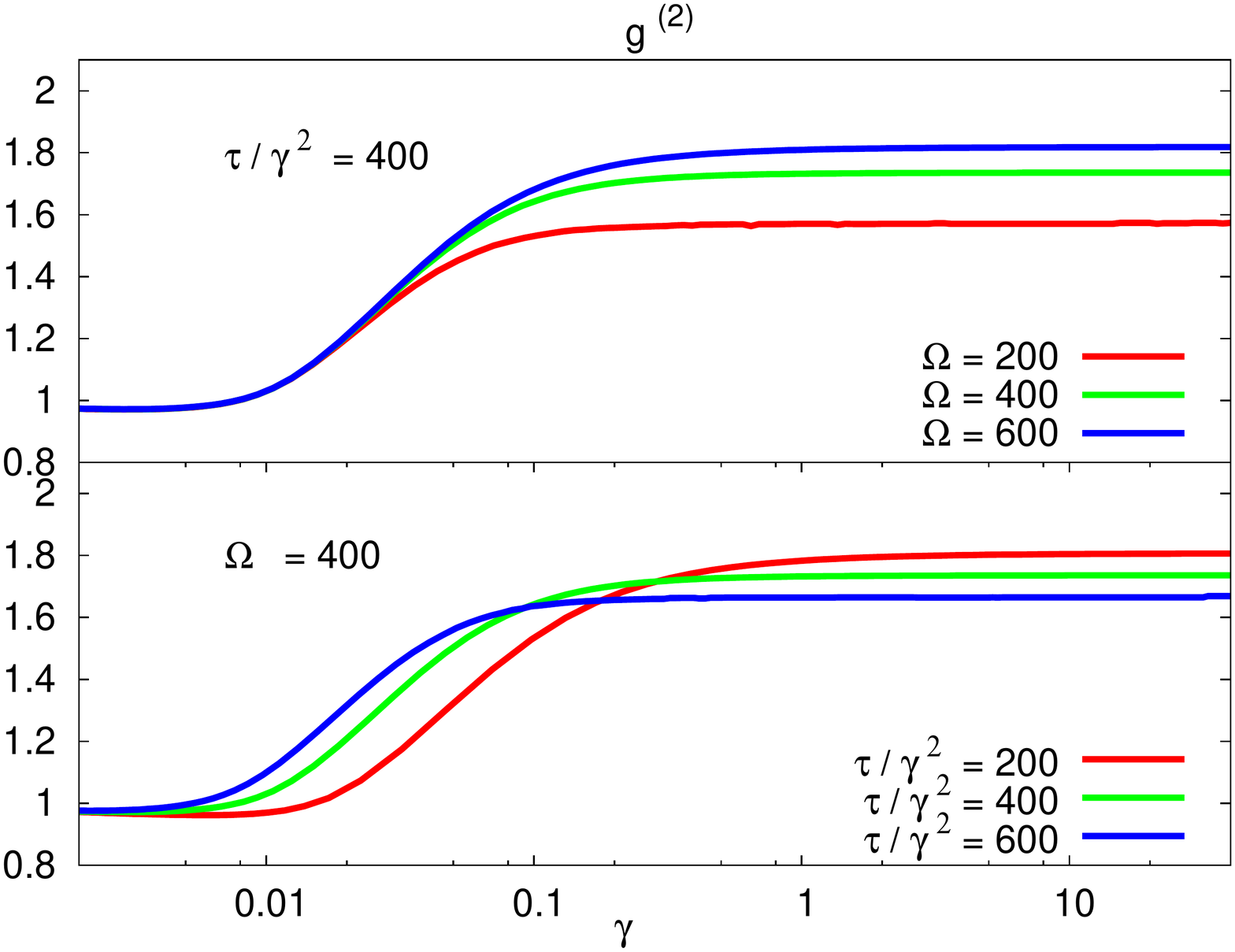}
    
       \caption{\label{fig:g2}The local pair correlation $g^{(2)}$ of the 2 component Bose gas
      as a function of the interaction strength $\gamma$ and for fixed values of $\Omega$ and
      for fixed ratio $\tau / \gamma^2$.}
\end{center}
\end{figure}
In this section we will present the results that don't belong to a limit regime. Furthermore in those parameter ranges, we give numerical results for the polarization of the 2CBG and the local pair correlator where neither the thermal, the charge nor the phase fluctuations dominate. They compete in the 2CBG state and therefore no perturbative approach but only the thermodynamic Bethe ansatz can predict a solution.
We will discuss and describe as much as possible the changes in
behaviour that occur in these intermediate regimes.
%We will first introduce this part with a discussion on the precision of the
%results.We then look at a curve in temperature to see explicitly the
%polarization effect of the Bose statistic. We also analyse a complete set of
%graphes on $\mu$ ($= \mu_1 + \mu_2$). And we will finally give a view of the
%population in an harmonic trap in function of the temperature and $\Omega$ ($=
%\mu_1 - \mu_2$).

%\subsection{Passing from a coupling regime from another}
The set of following graphs in figures \ref{fig:table_Pol} and \ref{fig:g2} show the behaviour of the two component Bose gas as a function of $\gamma$ and any fixed value of $\Omega$,$T$. The values of $\Omega$ and $T$ are chosen such that in each
case the ratio $\frac{\Omega}{T}$ goes from less to more than $1$. Following the qualitative description of \cite{GangardtPRL,KheruntsyanPRL} for the single-component case, the regimes of the gas are identified by the two dimensionless parameters $\gamma = \frac{c}{n_1+n_2}$ and $\tau=\frac{T}{(n_1+n_2)^2}$, respectively the interaction strength and the reduced temperature. As results presented hereafter are made for fixed interaction parameter ($c$), the ratio $\frac{\tau}{\gamma^2}=\frac{T}{c^2}$ is then constant and the regime is identified by the position on the $\gamma$ axis.
At the lowest value of $\gamma$, $\gamma \lesssim \tau \ll 1$ and the gas quasicondenses in a Gross-Pitaevskii (GP) regime with thermal fluctuations. At the other end, where $\gamma > 1$,  the regime is decoherent classical (DC) with $\tau \gg \max\{1, \gamma^2  \}$.
In the case of a quasicondensate, we see progessively the
ferromagntic effect with a completely polarized gas whereas the polarization reaches the value of an ideal paramagnetic gas when the 2CBG becomes DC.
From this simple view of the data, we can try to see how the temperature and
the relative chemical potential modify these phenomena.

The first column of figure \ref{fig:table_Pol} shows the effects of the
temperature on the polarization. In the region where $\gamma >1$ the linear density of each component is almost classical and the asymptotic value of the curves are given by %$\frac{n_1^0-n_2^0}{n_1^0+n_2^0}=
$\frac{e^{\beta \mu_1}-e^{\beta \mu_2}}{e^{\beta \mu_1}+e^{\beta \mu_2}}$ (see \ref{sec:DECOH}). Here the charge and coherent fluctuations are large and hence the statistics and the interaction of the gas don't play any role (the observables depend only on the temperature and chemical potentials).
In contrast, for $\gamma \ll 1$ the gas quasicondenses and the charge fluctuations vanish. The ratio $\tau / \gamma^2$ being large, the temperature fluctuations exceed the phase fluctuations and we see that $T$ doesn't influence the polarization much.

The second column of data shows the variations of the polarization as a function of $\Omega$.
The spontaneous ferromagnetism in the presence of the quasicondensate happens in high interaction strength when the relative chemical potential increases. In the YYT equations (\ref{eq:system}), the effect of $\Omega$ on the strings appears through the asymptotic value of the contribution of the
$n$-strings: $T \sum_{n = 1}^{\infty} \ln \left[ 1 + \exp (- 2 \frac{n
\Omega}{T}) \right]$. When $\Omega$ increases, the colour 2 spin-gapped
state effect are suppressed and the polarization resists higher charge fluctuation (higher $\gamma$).

Figure \ref{fig:g2} shows the local density-density correlation function as a function of $\gamma$ for different values of the relative chemical potential and temperature.
 On the top graph the ratio $\tau/\gamma^2$ is fixed; for $\gamma\ll 1$, the 2CBG is thus in a quasicondensate with important thermal fluctuations.
 In this regime the gas is ferromagnetic and $\Omega$ has no effect on the correlation.
 On the other hand, for large values of $\gamma$, the gas is DC and the asymptotic value of $g^{(2)}$ follows from Wick's theorem and the Boltzman distribution,
 $g_0^{(2)}=1+\frac{\sum_ie^{2\beta\mu_i}}{\left( \sum_ie^{\beta\mu_i} \right)^2}$. The first order corrections will be calculated later (\ref{sec:DECOH}).
 In the bottom figure, the curves for different temperatures show the nonmonotonic behaviour discussed in \cite{CauxPRA}.
 Close to the quasicondensate regime, the pair correlation increases with temperature: in the DC regime temperature has a destructive role on the correlation.

\section{Decoherent regimes}

\label{sec:DECOH}

                 \begin{center}
  \begin{figure}[h]
    \includegraphics[width=0.5\textwidth]{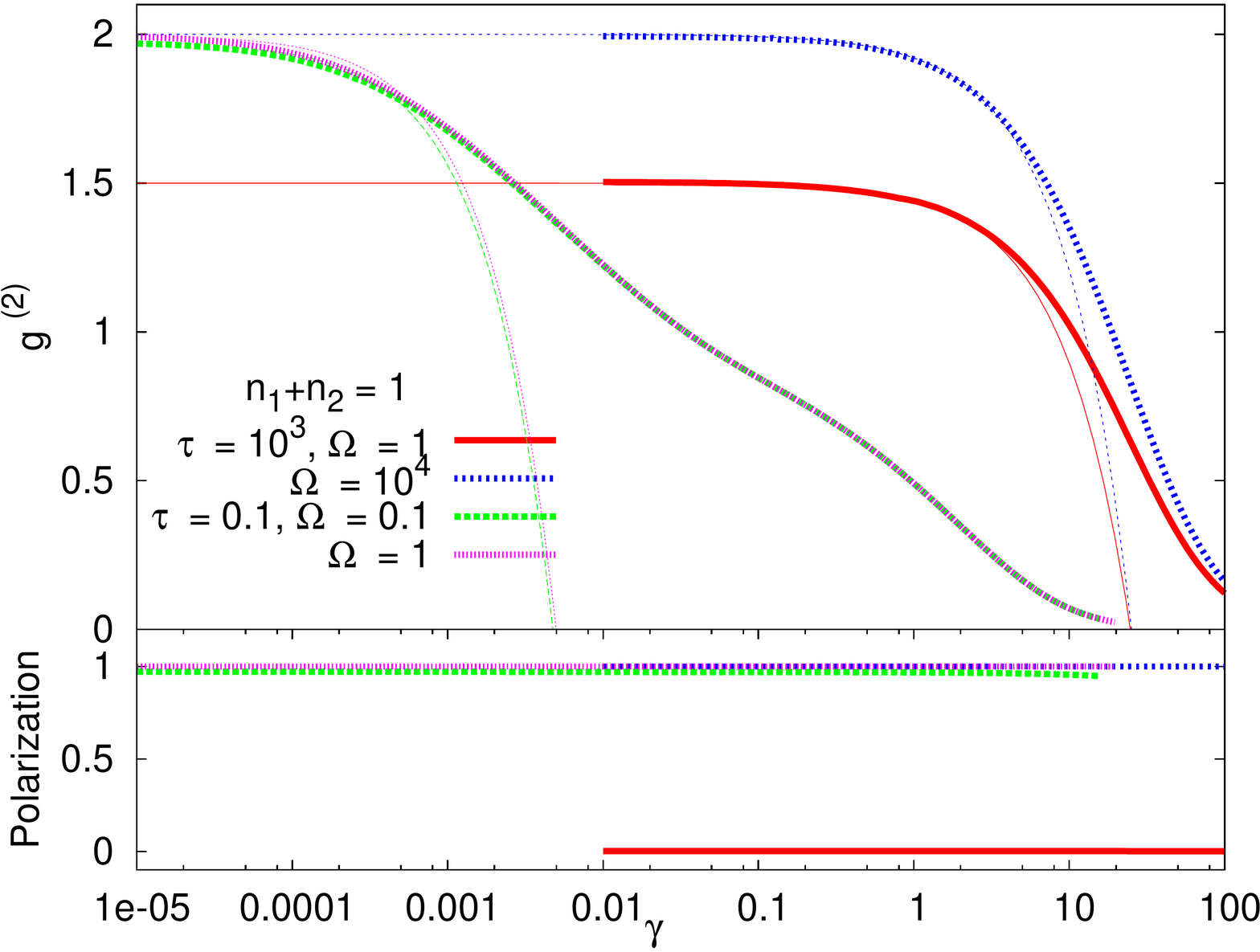}
    \caption{\label{fig:DC_DQ} Comparison between the numerical (thick lines) and the analytical results (thin lines)  in the decoherent quantum regime,
 $\sqrt{\gamma} \ll \tau \ll 1$ (\ref{eq:DQ}) and in the decoherent classical regime, $\tau \gg \max\{1,\gamma^2\}$ (\ref{eq:DC}).
The reduced temperature is fixed either to $\tau = 0.1$ with $\Omega = 0.1, 1$ for a decoherent quantum gas
 or to $\tau = 1000$ with $\Omega = 1, 10^4$ for a classical decoherent gas.
      At high value of the interaction strength when $\gamma \gg 1 $, the 2CBG enter a fermionization regime.}
  \end{figure}
\end{center}
In the limit of the weakly interacting Bose gas, ($\gamma \ll min \{ \tau^2,\sqrt{\tau} \}$) or in the high temperature regime ($\tau \ll max \{1, \gamma^2 \}$),
the phase and density fluctuations are large. Therefore one can notice that in the YYT equations \ref{eq:system}, the limit of either high temperature, $T^{-1}=\delta\ll 1$ with finite $c$, or low coupling $c=\delta\ll 1 $ with $T\neq0$, one recovers the thermodynamics of two ideal Bose gases up to $ \mathcal O(\delta^{2})$.
%In the system (\ref{eq:system}), we change the
%variable $k$ to $\bar{k} = \frac{k}{\sqrt{T}}$.
In this limit the convolutions of a function $g$ with the kernels described in equations (\ref{eq:system}) become:
\begin{eqnarray}
  a_n \ast g ( k) & %& \int^\infty_{-\infty} d \overline{ k'} \sqrt{T}  \frac{1}{\pi} \frac{n  c / 2}{(n c / 2)^2 + T ( \bar{ k} - \overline{ k'})^2} \cdot g ( \overline{ k'} \sqrt{T}) \nonumber\\
  = & \int^\infty_{-\infty} d \overline{ k'}  \underbrace{\frac{1}{\pi}
  \frac{n c / (2 \sqrt{T})}{(n c / (2 \sqrt{T}))^2 + ( \bar{ k} -
  \overline{ k'})^2}}_{\overset{\lim \delta \rightarrow
  0}{=} \delta ( \bar{ k} - \overline{ k'})} \cdot g ( \overline{ k'} \sqrt{T}) \nonumber\\
  & = & g ( \bar{ k} \sqrt{T}) \\
  &  &  \nonumber\\
  f \ast g ( k) & = & \int^\infty_{-\infty} d \overline{ k'} \underbrace{\frac{\sqrt{T}}{\cosh ( \frac{\pi}{c} \sqrt{T} (
  \bar{ k} - \overline{ k'})^{})}}_{\overset{\lim
  \delta \rightarrow 0}{=} c \delta ( \bar{ k} -
  \overline{ k'})} \cdot g ( \overline{ k'} \sqrt{T})
  \nonumber\\
  & = & c g ( \bar{ k} \sqrt{T}) 
\end{eqnarray}
Furthermore the Gibbs free energy resulting from this simplified system is:
\begin{eqnarray}
  \frac{G}{L} & = & - \frac{T}{2 \pi} \int^\infty_{-\infty} \ln \left[ 1 + e^ {-
  \varepsilon (k)/T} \right] d k \nonumber\\
  & = &\frac{T}{2 \pi} \int ^\infty_{-\infty} \ln \left[ \left( 1 - e^{( \mu_1 -
  k^2)/T} \right)\right] d k \nonumber\\
  & + & \frac{T}{2 \pi} \int ^\infty_{-\infty} \ln \left[ \left( 1 - e^{ (\mu_2 - k^2)/T} \right)
  \right] d k 
\end{eqnarray}
which is the sum of the Gibbs energy of two ideal Bose gases.

First order corrections can then be effectively described using perturbation theory and the reduced temperature, $\tau=\frac{T}{(n_1 + n_2)^2}$, allows one to distinguish between the decoherent quantum regime (DQ) for $\sqrt{\gamma} \ll \tau \ll 1 $ and the decoherent classical (DC) regime with
 $\tau \gg \max\{1,\gamma^2\}$ \cite{GangardtPRL,KheruntsyanPRL,Deuar_PRA_2009}. 

 We use Feynman diagrams to express the perturbed Gibbs free energy. An explicit expression is then calculated for the local pair correlation function,
 $g^{(2)}$ in the two decoherent regimes (DQ \& DC) to first order.

\begin{figure*}[]
%\begin{widetext}
  %\includegraphics[scale=0.5]{/Users/Pro/eqbose/MPI/data/T/2CBG_T.pdf}
  \includegraphics[width=\textwidth]{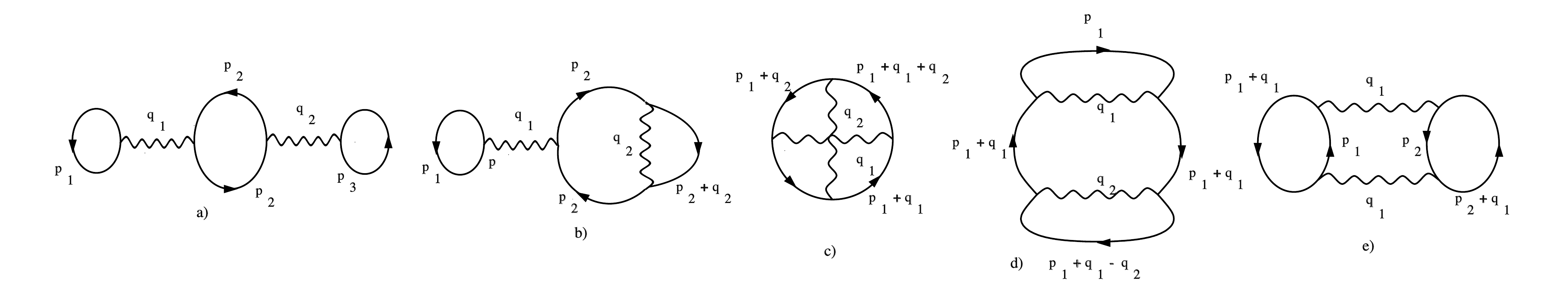}
  \caption{Connected Feynman diagrams in second order perturbation of interaction. Labels are related to terms of eq. (\ref{eq:G^2}) }
\label{fig:diagrams}
%\end{widetext}
\end{figure*}
The partition function of the 2 component 1D Bose gas in the Feynman path integral formalism is
\begin{eqnarray}
{\mathcal Z} &=&\int {\mathcal D}{ ( \bar{\mathbf \Psi},\mathbf \Psi )} e^{-S[\bar{\mathbf \Psi},\mathbf \Psi]}\nonumber\\
 S[  \bar{ \mathbf \Psi}, \mathbf \Psi] &=&\int^{\beta}_0d\tau \int d r \sum_a\bar{\Psi}_a \partial_\tau \Psi_a -\mathcal H(\bar{ \mathbf \Psi}, \mathbf \Psi)
\end{eqnarray}
where ${\mathbf \Psi}(r,\tau)$ is a space and imaginary time-dependent spin-$1/2$ field and $\mathcal H$ is the Hamiltonian density from (\ref{eq:H_N}).
At first order, the correction to the Gibbs free energy following from Wick's theorem is $G^{(1)} =2c \left[ {n^0}^2 - n^0_0 n^0_1  \right] + {\mathcal O}(c^2) $
%\begin{eqnarray}
%G &=&2c \left[ \left( \sum_{a}\int d\sigma dr G_a(r,\sigma) \right) \right. ^2 \nonumber\\
%&-&   \left. \int d\sigma d r G_0(r,\sigma) G_1(r,\sigma) \right] + \mathcal O(c^2)  
%G^{(1)} &=&2c \left[ \left( 1/L\sum_{a,k,m} G_a(k,\omega_m) \right) \right. ^2 \nonumber\\
%#&-&   \left. 1/L \sum_{k,m} G_0(k,\omega_m) \sum_{k',n}G_1(k',\omega_n) \right] + \mathcal O(c^2)  
%G^{(1)} &=&2c \left[ \left( \sum_{a} n^0_a \right)^2 - n^0_0 n^0_1  \right] + {\mathcal O}(c^2)  
%\end{eqnarray}
%with the Green's function in momentum and Matsubara frequencies: $G_a(k,\omega_n)=\frac{1}{i\hbar \omega_n-\hbar^2k^2/2m+\mu_a}$
with the free linear density of the $a$-th component: $ n^0_a=\frac{T}{L}\sum_{k,n}\frac{1}{i\hbar \omega_n-\hbar^2k^2/2m+\mu_a}$ and the total free linear density: $n^0=  \sum_a n^0_a $.
For the second order in $c$, the diagrammatic representation gives five contributions shown in figure \ref{fig:diagrams} that give the free energy density corrections:
%\begin{widetext}
\begin{widetext}
\begin{eqnarray}
%G &=&2c \left[ \left( \sum_{a}\int d\sigma dr G_a(r,\sigma) \right) \right. ^2 \nonumber\\
%&-&   \left. \int d\sigma d r G_0(r,\sigma) G_1(r,\sigma) \right] + \mathcal O(c^2)  
%G^{(2)} &=&-\frac{c^2 }{2 }\frac{T^3}{L^3} \left[ 8\left( \sum_{a,k,m} G_a(k,\omega_m) \right)\sum_b \sum_{k,m} G_b(k,\omega_m) \sum_{k',n} G_b^2(k',\omega_n) \right.  \nonumber\\
%&+& 4\left( \sum_{a,k,m} G_a(k,\omega_m) \right)^2 \sum_{b,k',n}G_b^2(k',\omega_n)  + 4\sum_a\left( \sum_{k,m} G_a(k,\omega_m) \right)^2 \sum_{k',n}G_a^2(k',\omega_n)\nonumber\\
%& +& \left.  2 \sum_{k,m,a}2 \left( \left( \sum_{k',n} G_a(k',\omega_n)G_a(k + k',\omega_m+\omega_n) \right)^2 + \sum_{k',n} G_a(k',\omega_n)G_a(k + k',\omega_m+\omega_n) \sum_{k'',o} G_{|a-1|}(k'',\omega_o)G_{|a-1|}(k + k'',\omega_m+\omega_o) \right) \right]
G^{(2)} &=&-\frac{c^2 }{2 } \left[ 8\overbrace{ n^0 \sum_bn^0_b\partial_{\mu_b}n^0_b}^{b)} + 4 \overbrace{ {n^0}^2 \sum_{b}\partial_{\mu_b}n^0_b}^{a)}   
+   4 \underbrace{\sum_a(n^0_a)^2 \partial_{\mu_a}n^0_a}_{d)} +  2 \sum_{a=0,1}\left( \underbrace{P_{a,a}}_{c)}+\underbrace{ P_{a,a} +P_{a,|a-1|} }_{e)} \right) \right]  
 + \mathcal O(c^3) \hspace{0.7cm}
\label{eq:G^2}
\end{eqnarray}
\end{widetext}
where the first three terms correspond to the diagrams $b)$, $a)$ and $d)$ and where $c)$ and $e)$ provide the last terms containing the double polarization bubbles which are defined as
% \begin{widetext}
\begin{widetext}
\begin{eqnarray}
  P_{a,b}&=&\sum_{m} \int^\infty_{-\infty} d k \: \left ( \sum_{n} \int^\infty_{-\infty} d l  \: G_{a,m+n}(k+l)G_{a,n}(l) \sum_{n'} \int^\infty_{-\infty} d l'\:  G_{b,m+n'}(k+l')G_{b,n'}(l')\right )
\end{eqnarray}
\end{widetext}
%\end{widetext}
with the Green function $G_{a,n}(l)=\frac{1}{i\hbar \omega_n- \hbar^2l^2/2m+\mu_a}$. The local pair correlation results from equation (\ref{eq:g^2}) and an analytic expression as a function of $c,T,\mu_i$ is given in the two decoherent regimes.
In the DQ regime, $\sqrt{\gamma} \ll \tau \ll 1 $ and $\mu_1,\mu_2 \ll T$, by taking the leading order in the Bose occupation number,
 the free linear density is $n^0_a=\frac{T}{2\sqrt{-\mu_a}} $ and the double polarization bubble is $P_{a,b}=\frac{n^0_a+n^0_b-({n^0_a}^2+{n^0_b}^2)/(n^0_a+n^0_b)}{\tau_a\tau_b}$.
 For a compact notation we define the $a$th component reduced temperature by $\tau_a=\frac{T}{{n^0_a}^2}$.
 We find so the local pair correlation to be
\begin{widetext}
\begin{eqnarray}
g^{(2)} & =&\frac{{  n^0 }^2 + \sum_i {n^0_i}^2}{{n^0}^2}   +4 \gamma\left[ \frac{2}{\tau_1 \tau_2} - \frac{1}{ {n^0}^2}(\frac{n^0_1}{\tau_1^2}+\frac{n^0_2}{\tau_2^2}) \right]
- 4\gamma \frac{{n^0}^2 - \sum_i {n^0_i}^2}{{n^0}^2} \left[  4( \frac{1}{\tau^2_1} + \frac{1}{\tau^2_2}) +  \frac{1}{\tau_1 \tau_2} \right]
 + {\mathcal O}(\gamma^2)  
\label{eq:DQ}
\end{eqnarray}
\end{widetext}
In the DC regime, $\tau \gg \max\{1,\gamma^2\}$ and $\mu_1,\mu_2 \gg T$, the bosonic occupation number becomes the Boltzmann distribution and $n^0_a=\sqrt{\frac{\pi}{\beta}}\frac{ e^{\beta\mu_a}}{2}$, $P_{a,b} = n^0_a n^0_b \sqrt{\frac{\pi\beta}{8}}$. The pair correlator becomes
\begin{widetext}
\begin{eqnarray}
g^{(2)} & =&\frac{{n^0}^2 + \sum_i {n^0_i}^2}{{n^0}^2} 
  - \sqrt{\frac{\pi}{2}} \gamma \left( \frac{1}{\sqrt{\tau_1}} +\frac{1}{\sqrt{\tau_2}} \right)\left[ \frac{{n^0}^2 + \sum_i {n^0_i}^2}{{n^0}^2} \right]
+ \frac{4 \gamma}{\sqrt{\tau_1 \tau_2}}\left[ 1-2 \frac{{n^0}^2 - \sum_i {n^0_i}^2}{{n^0}^2}\right] + {\mathcal O}(\gamma^2) 
\label{eq:DC}
\end{eqnarray}
\end{widetext}
In order to illustrate this result, we compare the value of $g^{(2)}$ computed to the first order in $\gamma$ (eq.\ref{eq:DQ} and \ref{eq:DC}) with the numerical results using
 (\ref{eq:g^2}) in figure \ref{fig:DC_DQ}. The curves calculated at fixed particle density and reduced temperature show that the numerical results follow nicely
 the analytical expansion until either the thermal fluctuations become too strong for the DQ gas ($\gamma\sim 10^{-3}$)
 or the charge fluctuations become important in the DC regime when $\gamma \sim 10$. As the interaction strength increases, we progressively switch to a high temperature
 Tonks-Girardeau like fermionization regime for the DC curves and to a ferromagnetic fermionization for the DQ case.
It would be interesting as well to compare the results for a DQ gas in very low $\Omega$ such that the polarization is low and $g^{(2)}$ reaches the value $3/2$
 but this implies a calculation for a very high number of functions, $n_{max}$ with a large number of points.
We couldn't afford then the number of iterations necessary to have a converged solution.
\section{Tonks-Girardeau regime}
\label{sec:TONKS}
In the extreme case of impenetrable particles ($\gamma \rightarrow \infty$),
M. Girardeau
\cite{GirardeauJMP1} showed the correspondence between impenetrable  Bose and Fermi wave functions. While the statistics of the bosons wave function remains symmetric, there is no more overlap between the neighbor particles. 
In the case of 2CBG, the charge part of the wave function behave like a one-component free fermion gas and noninteracting distinguishable spin$-1/2$ since any spin-spin exchange vanishes \cite{DeuretzbacherPRL100,GuanPRA76,TakahashiBOOK}.
In both 1CBG and 2CBG, the local density-density correlation function then naturally vanishes since there is no double space occupancy.

\begin{center}
  \begin{figure}[h]
    \includegraphics[width=0.5\textwidth]{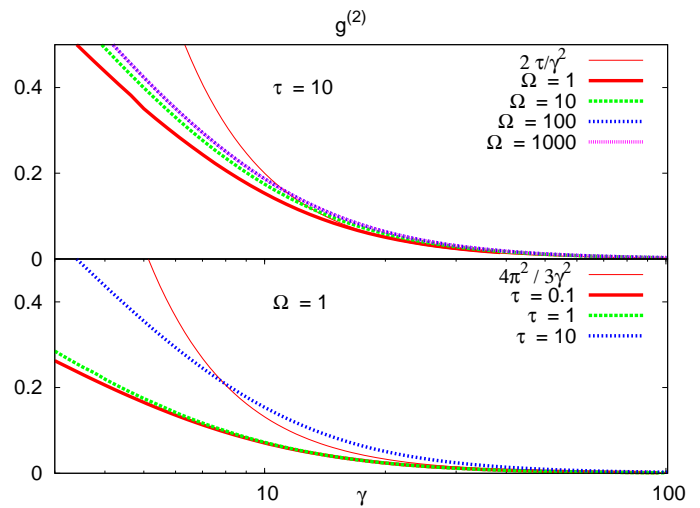}	
    \caption{\label{fig:TG} In the Tonks-Girardeau regime, the pair-pair correlation decay to zero when the correlation strength becomes large ($\gamma\gg \max( 1,\sqrt{\tau})$).
 The top graph shows the high-temperature fermionization and its dependence in $\Omega$. 
The 1CBG asymptotic behavior in $\tau/\gamma^2$ is also shown in comparison with the fully polarized 2CBG ($\Omega\gg T$).
In the bottom graph we show curves with different reduced temperatures that
 we compare with the analytical expression at zero temperature \cite{GuanPRA76}. The 2CBG is here ferromagnetic and the value of $g(2)$ doesn't depend on the relative chemical potential.}
  \end{figure}
\end{center}

In the strong coupling regime ($\gamma \gg \max( 1,\sqrt{\tau})$) with quantum degeneracy ($\tau\ll 1$),
 the finite-temperature corrections are markedly different in a Lieb-Liniger gas \cite{KheruntsyanPRL} and the spinor Bose gas \cite{GuanPRA76}  with a different exponent.
In figure \ref{fig:TG}, bottom part, we represent this analytical result at zero temperature (thin line) next to numerical results with decreasing temperature for fixed density of particles.
 We observe that for $\gamma \gg 10$, the value of $g^{(2)}$ decreases with $\tau$ and converges to this $T=0$ analytical result where the 2CBG is ferromagnetic and doesn't depend on $\Omega$.
For a high-temperature fermionization ($\gamma^2\gg \tau \gg 1$), Kheruntsyan et al. \cite{KheruntsyanPRL} give the first order correction in $\tau/\gamma^2$ for $g^{(2)}$ for a Lieb-Liniger gas.
However the approach of free fermions with a $1/\gamma$ perturbation is not applicable in the two component case, therefore the correction to the fermionization regime are unknown.
In figure \ref{fig:TG}, top part, we show next to the 1 component asymptotic curve, the decay of $g^{(2)}$  for a fixed reduced temperature and various relative chemical potential.
 As $\Omega$ reaches $100$, the 2CBG polarization is saturated and the correlator decays like a Lieb-Liniger gas.
 
\section{Conclusion}
In conclusion, we have studied the equilibrium thermodynamic properties of exactly solvable interacting one-dimensional two-component Bose gas systems as a function
 of their external canonical or grand canonical parameters
 (either temperature, interaction strength and total and relative chemical potential or temperature, interaction strength, densiy of particle and relative chemical potential).
Our method was based on the solution of thermodynamic Bethe ansatz equations and yields quantitative predictions which should be experimentally accessible using cold atomic systems.
We particularly would like to clarify that solving the non linear integrable equations is possible with a very good control of numerical precision.

{\it Note:} as our manuscript was being completed, a different but equivalent set of equations was proposed in \cite{Klumper1103.6152}.
 While this set of equations is at first sight more economical, we find and demonstrate here that the solution of the infinite set of TBA equations is feasible and practical, robust and reliable.
 The TBA dressed energies in \ref{eq:system} are relatively smooth functions of a real variable, while the functions of \cite{Klumper1103.6152} are are of a complex variable.
 The computational effect required by the two methods are thus probably comparable. On the other hand, the fact that results from this alternate method coincide with our results here
 (and our earlier summary \cite{CauxPRA}) interestingly confirms that the string hypothesis can be trusted when computing equilibrium thermodynamic results,
 as expected from general arguments based on the structure of the Bethe equations \cite{TsvelickAP32}.

\begin{appendix}

\section{Thermodynamics from Bethe Ansatz}
\label{sec:TBA}
We model the system of 2 component bosons with SU(2) bosonic fields evolving in a 1 dimensional
continuum space of length $L$ with a delta-function interaction. The
Hamiltonian is then:
\begin{eqnarray}
%  H = \int_0^L d x \sum_{a \epsilon \{- 1, 1\}} \left\{ \partial_x
%  \Psi^{\dagger}_a (x) \partial_x \Psi_a (x) + c \int_0^L d x' \delta (x - x')
%  \sum_{b \epsilon \{- 1, 1\}} \Psi^{\dagger}_a (x) \Psi^{\dagger}_b (x')
%  \Psi_a (x) \Psi_b (x') \right\}
  H = \int_0^L d x \sum_{a \in \{- 1, 1\}} \partial_x
  \Psi^{\dagger}_a (x) \partial_x \Psi_a (x)  \nonumber \\
 + c 
  \sum_{b,a \in \{- 1, 1\}} \Psi^{\dagger}_a (x) \Psi^{\dagger}_b (x)
  \Psi_b (x) \Psi_a (x) 
\end{eqnarray}

With $c = \frac{g_{\text{1D}} \cdot m}{\hbar^2}$, $g_{\text{1D}}$ the 1D coupling constant
and $m$ the mass of the bosons. 

Integrating the string structure,$ \lambda^n_{\alpha,j} = \Lambda^n_{\alpha} + \frac{\mathi c}{2} (n + 1 - 2 j),\:j = 1,\ldots, n $, in the scattering equations in (\ref{eq:Bethe_eq}) and defining $e_n (\lambda) = \frac{\lambda -  \mathi c  n / 2}{\lambda + \mathi c  n / 2}$, the scattering equations become:

\begin{widetext}

\begin{eqnarray}
  e^{\mathi k_j L} & = & - \prod_{l = 1}^N \frac{k_j - k_l + \tmop{ic}}{k_j -
  k_l - \tmop{ic}} \prod^{\infty}_{n = 1} \prod^{N_n}_{\alpha = 1} \frac{k_j -
  \Lambda^n_{\alpha} - \frac{\mathi n c}{2}}{k_j - \Lambda^n_{\alpha} +
  \frac{\mathi n c}{2}} \nonumber\\
  _{}^{} \prod_{p = 1}^N \frac{k_p - \Lambda^n_{\alpha} + \frac{\mathi n
  c}{2}}{k_p - \Lambda^n_{\alpha} - \frac{\mathi n c}{2}} & = & (- 1)^n
  \prod^{}_{m, \beta} \left\{ \begin{array}{l}
    e^2_2 (\Lambda) e^2_4 (\Lambda) e^2_{n - m + 4} (\Lambda) \ldots e^2_{2 n
    - 2} (\Lambda) e^{}_{2 n} (\Lambda), \hspace*{\fill} m = n\\
    e_{n - m} (\Lambda) e^2_{n - m + 2} (\Lambda) e^2_{n - m + 4} (\Lambda)
    \ldots e^2_{n + m - 2} (\Lambda) e^{}_{n + m} (\Lambda), m \neq n
  \end{array} \right. \nonumber\\
  &  & 
  \label{string_Bethe_eq}
\end{eqnarray}

with the notation: $ \Lambda =
\Lambda^n_{\alpha} - \Lambda^m_{\beta} .$

In logarithm form, with $\frac{1}{i} \ln (e_n (\Lambda)) = \phi_n (\Lambda) = - \pi +2
\tmop{atan} ( \frac{\Lambda}{c n / 2})$, we have:
\begin{eqnarray}
  k_j & = & 2 \pi \frac{I_j}{L} - \frac{1}{L} \sum_{l =
  1}^N \phi_2 (k_j - k_l) + \frac{1}{L} \sum^{\infty}_{n = 1}
  \sum^{N_n}_{\alpha = 1} \phi_n (k_j - \Lambda_{\alpha}^n) \nonumber\\
  \frac{1}{L} \sum_{p = 1}^N \phi_n (k_p - \Lambda_{\alpha}^n) & = &
  \frac{1}{L} 2 \pi J^n_{\alpha} \nonumber\\
  & + & \frac{1}{L} \sum^{\infty}_{m = 1} \sum^{N_m}_{\beta = 1} \left\{
  \begin{array}{l}
    2 \phi^{}_2 (\Lambda_{\alpha}^n - \Lambda^m_{\beta}) + 2 \phi^{}_4
    (\Lambda_{\alpha}^n - \Lambda^m_{\beta}) \ldots \phi_{2 n}
    (\Lambda_{\alpha}^n - \Lambda^m_{\beta}), \hspace*{\fill} m = n\\
    \phi_{|n - m|} (\Lambda_{\alpha}^n - \Lambda^m_{\beta}) + 2 \phi_{|n - m|
    + 2} (\Lambda_{\alpha}^n - \Lambda^m_{\beta}) \ldots \phi^{}_{n + m}
    (\Lambda_{\alpha}^n - \Lambda^m_{\beta}), m \neq n
  \end{array} \right. \nonumber\\
  &  & 
\label{log_Bethe_eq}
\end{eqnarray}

\end{widetext}
$\{I_j \}$ is a set of $N$ numbers in $\mathbbm{Z}+
\frac{1}{2}$ and $\{J_{\alpha}^n \}$ are $M_n$ sets of $n$ numbers in
$\mathbbm{Z}$ ($\mathbbm{Z}+ \frac{1}{2}$) if $M_n$ is even (odd).
These Bethe equations map the sets $\{I_j \}, \{J^n_{\alpha} \}$ to the set of
rapidities and isospin-rapidities, $\{k_j \}$ and$ \{\Lambda^n_{\alpha}
\}$.

\subsection{Thermodynamic limit}
In the limit $N, L \rightarrow \infty$ with the ratio $\frac{N}{L}$ kept constant,
the sets of rapidities ($\{k_j \}$ and $\{\Lambda^n_{\alpha} \}$) and quantum numbers ($\{I_j \}$ and $ \{J^n_{\alpha} \}$)
 are replaced by continuous functions of particle root densities in real parameter space:

\begin{eqnarray}{}
  \rho (x) = \frac{1}{L} \sum_j \delta (x - \frac{I_j}{L}), \rho (k') =
  \frac{1}{L} \sum_j \delta (k- k_j ( \frac{I_j}{L}))\nonumber \\
  \sigma^n (y^{}) = \frac{1}{L} \sum_j \delta (y^{} - \frac{J^n_\alpha}{L}) ,  \hspace{3cm}\nonumber\\
   \sigma^n (\Lambda') = \frac{1}{L} \sum_j \delta (\Lambda - \Lambda_\alpha^n
  ( \frac{J^n_\alpha}{L}), \forall n \hspace{.5cm}. \hspace{1.5cm}
\end{eqnarray}
Hole densities, $\rho_h, \sigma^n_h$ are similarly defined from the complementary sets $\{
\tilde{I}_i \}, \{ \tilde{J}^n_{\alpha} \}$, and the total root densities are $\rho_t(k)=\rho_h(k)+\rho_h(k)$ and $\sigma_t^n (\Lambda)=\sigma_h^n (\Lambda)+\sigma^n (\Lambda) $.
The thermodynamic limit allows one to replace the discrete sum by an integral over a continuum.
\begin{eqnarray}
  \text{$\rho_t (x) = \frac{1}{L} \sum_{I \epsilon \{ \tilde{I}_i \}, \{I_i
  \}} \delta (x - \frac{I}{L})$}  \underset{\text{Th.L.}}{\longrightarrow} 
  \int^{\infty}_{- \infty} d x' \delta (x - x') = 1 \nonumber\\
\sigma_t^n (y)  \underset{\text{Th.L.}}{\longrightarrow}  1 \hspace{3cm} 
\end{eqnarray}
and the indexations of the rapidities by the quantum numbers, ($k_j ( \frac{I_j}{L}), \Lambda_\alpha^n
  ( \frac{J^n_\alpha}{L})$) become continuous functions :  $k (x)$ and $\Lambda^n (y)$.
An important point of the thermodynamic limit is the assumption that all
the density functions are in $\mathcal{C}^{\infty}$. However some sets of $\{I_j \}$, $\{J^n_{\alpha} \}$ that are solutions of (\ref{string_Bethe_eq}),
could provide no differentiable functions. For instance if all the rapidities are
grouped in a block Fermi-sea like. But the role of these solutions play a negligible
role in the thermodynamic limit due to the fact that their weight in the set
of all solutions goes to zero. Physically, they represent the solutions with
low entropy. Finally, under the thermodynamic limit (\ref{log_Bethe_eq}) becomes:
\begin{widetext}
\begin{eqnarray}
  ^{} k_{} (x) & = & 2 \pi x - \int^\infty_{-\infty} \phi_2 (k_{} - k_{}') \rho (k') d k' +
  \sum^{\infty}_{m = 1} \int^\infty_{-\infty} \phi_m (k_{} - \Lambda_{}) \sigma^m (\Lambda) d \Lambda \nonumber\\
  \int^\infty_{-\infty} \phi_n (k_{} - \Lambda_{}) \rho (k) d k & = & 2 \pi y \nonumber\\
  & + & \sum^{\infty}_{m = 1} \int^\infty_{-\infty} \sigma^m (\Lambda') d \Lambda' \left\{
  \begin{array}{l l}
    2 \phi^{}_2 (\Lambda_{} - \Lambda') + \ldots \phi_{2 n} (\Lambda_{} -
    \Lambda'),  & m = n\\
    \phi_{|n - m|} (\Lambda_{} - \Lambda'_{}) + \ldots \phi^{}_{n + m}
    (\Lambda_{} - \Lambda'), \: &   m \neq n
  \end{array} \right. \hspace{0.5cm}.\nonumber\\
  &  & 
\end{eqnarray}
\end{widetext}
\subsection {YYT equations}
 Following the method of C. N. Yang and C. P. Yang \cite{yang&yang}, the equilibrium state is determined by minimization of the Gibbs free energy in the grand
canonical ensemble. With $G$ the Gibbs free energy, $E$ the internal energy, $S$ the entropy, we have:
\begin{widetext}\begin{eqnarray}
  G & = & E - T S - \mu_1 N_1 - \mu_2 N_2 \nonumber\\
  \frac{E}{L} & = & \int^\infty_{-\infty} d k \rho (k) k^2 \nonumber\\
  \frac{S}{L} & = & \int^\infty_{-\infty} d k \left[ (\rho + \rho_h) \ln (\rho + \rho_h) - \rho
  \ln (\rho) - \rho_h \ln (\rho_h) \right] + \sum^{\infty}_{n = 0} \int^\infty_{-\infty} dk \left[
  (\sigma^n + \sigma^n_h) \ln (\sigma^n + \sigma^n_h) - \sigma^n \ln
  (\sigma^n) - \sigma^n_h \ln (\sigma^n_h) \right] \nonumber\\
  \mu_1 n_1 &+& \mu_2 n_2  =  \int^\infty_{-\infty} d k \Omega \left( \rho - 2 \sum_n n
  \sigma^n \right) + \mu \rho 
\end{eqnarray}
\end{widetext}
with $L$ the length of our system, $n_i = \frac{N_i}{L}$ the density of $i$th
component particles and $\mu =\frac{ \mu_1 + \mu_2}{2}, \Omega = \frac{\mu_1 - \mu_2}{2}$. The
condition of equilibrium is then:
\begin{widetext}\begin{eqnarray}
 \delta \rho \frac{\partial G}{\partial \rho} + \delta \rho_h  \frac{\partial
  G}{\partial \rho_h} + \sum_n \delta \sigma^n  \frac{\partial G}{\partial
  \sigma^n} + \delta \sigma^n_h  \frac{\partial G}{\partial \sigma_h^n} & = &
 0  \: \bigg|_{\rho,  \sigma^n \text{solution of BE} }
\end{eqnarray}
\label{dG}from which one derives:
\begin{eqnarray}
  \varepsilon (k) & = & k^2 - \mu - \Omega - T \cdot a_2 \ast \ln (1 + e^{-
  \varepsilon / T}) - \sum_n T \cdot a_n \ast \ln (1 + e^{- \varepsilon_n /
  T}) \nonumber\\
  \varepsilon_n (k) & = & 2 n \Omega + T \cdot a_n \ast \ln (1 + e^{-
  \varepsilon_{} / T}) + T \cdot \sum_m T_{m n} \ast \ln (1 + e^{-
  \varepsilon_m / T}),\: n =1,2,\ldots 
\end{eqnarray}
\label{Thermo_1st} with
\begin{eqnarray}
  T_{n m} (\Lambda^{}) & = & \left\{ \begin{array}{l}
    2 a^{}_2 (\Lambda_{}) + 2 a^{}_4 (\Lambda_{}^{}) \ldots a_{2 n}
    (\Lambda_{}), \hspace*{ 1cm} m = n\\
    a_{|n - m|} (\Lambda_{}^{}) + 2 a_{|n - m| + 2} (\Lambda_{}) \ldots
    a^{}_{n + m} (\Lambda), m \neq n
  \end{array} \right.\\
  \varepsilon (k) & = & T \ln ( \frac{\rho_h (k)}{\rho_{} (k)})\\
  \varepsilon_n (k) & = & T \ln ( \frac{\sigma_h^n (k)}{\sigma^n_{} (k)})
\end{eqnarray}\end{widetext}
%It have been shown that $\varepsilon (k)$ is the particle excitation energy
%([Yang \& Yang 1969]) and we can phenomenologicaly think about $\varepsilon_n
%(k)$ as the bound state energy of $n$ bosons.
The term $T_{\tmop{mn}}$ which implies a coupling between every
$\varepsilon_m (k)$ would severely slow down any numerical solving. But following the development of M. Takahashi
\cite{TakahashiPTP,TakahashiBOOK}, the system is partially decoupled  and this term disappears:

\begin{widetext}\begin{eqnarray}
  \varepsilon (k) & = & k^2 - \mu - \Omega - T \cdot \left( a_2 \ast \ln
  \left[ 1 + \exp (- \frac{\varepsilon}{T}) \right] \right) (k) - T \sum_{n =
  1}^{\infty} \left( a_n \ast \ln \left[ 1 + \exp (- \frac{\varepsilon_n}{T})
  \right] \right) (k) \\
  &  &  \nonumber\\
  \varepsilon_n (k) & = & \frac{T}{2 c} \left\{ f \ast \left( \ln
  \left[ 1 + \exp ( \frac{\varepsilon_{n + 1}}{T}) \right] + \ln \left[ 1 +
  \exp ( \frac{\varepsilon_{n - 1}}{T}) \right] \right) \right\} (k),
  \hspace*{\fill} (n \neq 1) \\
  &  &  \nonumber\\
  \varepsilon_1 (k) & = & \frac{T}{2 c} \left\{ f \ast \left( \ln
  \left[ 1 + \exp ( \frac{\varepsilon_2}{T}) \right] + \ln \left[ 1 + \exp (-
  \frac{\varepsilon_{}}{T}) \right] \right) \right\} (k) 
  \label{eq:system_appendix}
\end{eqnarray}\end{widetext}
with the convolution notation: $(f \ast g) (k) = \int f (k - k')
g (k') d k'$, $a_n (k) = \frac{1}{\pi} \frac{n c / 2}{(n c / 2)^2 + k^2}$, 
$f (k) = 1 / \cosh ( \frac{\pi}{c} k)$.

We can easly calculate the two asymptotic limit similarly to the results for the istropic spin chain of
M.Takahashi \cite{TakahashiBOOK}:
\begin{widetext}\begin{eqnarray}
  \lim_{n \rightarrow \infty} \frac{\varepsilon_n (k)}{n} & = & 2 \Omega \\
  \lim_{k \rightarrow \infty} \varepsilon_n (k) (\equiv
  \varepsilon_n^{\infty}) & = & 2 \Omega n + T \cdot \ln \left[ \left( \frac{1
  - \exp (- \frac{2 \Omega}{T} (n + 1))}{1 - \exp (- \frac{2 \Omega}{T})}
  \right)^2 - \exp (- \frac{2 \Omega}{T} n) \right]
  \label{eq:limits}
\end{eqnarray}
\section{Dressed energy derivatives}
The derivatives of $\varepsilon (k)$ and $\varepsilon_n (k)$
are useful for calculation of free energy derivatives. Differentiating
(\ref{eq:system}) and (\ref{eq:limits}) by $\nu =  \mu, \Omega $, we get:

\begin{eqnarray}\frac{\partial \varepsilon}{\partial \nu} (k) & = & - 1 + \left( a_2 \ast
  \frac{\partial \varepsilon / \partial \nu}{1 + \exp (
  \frac{\varepsilon}{T})} \right) (k) + \sum_{n = 1}^{\infty} \left( a_n \ast
  \frac{\partial \varepsilon_n / \partial \nu}{1 + \exp (
  \frac{\varepsilon_n}{T})} \right) (k) \nonumber\\
  \frac{\partial \varepsilon_n}{\partial \nu} (k) & = & \frac{1}{2 c} \left[
  f \ast \left(  \frac{\partial \varepsilon_{n + 1} / \partial
  \mu}{1 + \exp (- \frac{\varepsilon_{n + 1}}{T})} + \frac{\partial
  \varepsilon_{n - 1} / \partial \nu}{1 + \exp (- \frac{\varepsilon_{n -
  1}}{T})} \right) \right] (k), \hspace*{\fill} (n \neq 1) \nonumber\\
  \frac{\partial \varepsilon_1}{\partial \nu} (k) & = & \frac{1}{2 c} \left[
  f \ast \left(  \frac{\partial \varepsilon_2 / \partial \nu}{1 +
  \exp (- \frac{\varepsilon_2}{T})} - \frac{\partial \varepsilon_{} / \partial
  \mu}{1 + \exp ( \frac{\varepsilon_{}}{T})} \right) \right] (k) \nonumber\\
    \lim_{n \rightarrow \infty} \frac{\partial \varepsilon_n / \partial \Omega
  (k)}{n} & = & 2 \nonumber\\
  \frac{\partial \varepsilon_n^{\infty}}{\partial \Omega} & = & \frac{2 \left(
  1 - \exp (- \frac{2 \Omega}{T} (n + 1)) \right)}{\left( 1 - \exp (- \frac{2
  \Omega}{T}) \right) \left( 1 - \exp (- \frac{2 \Omega}{T} n) \right) \left(
  1 - \exp (- \frac{2 \Omega}{T} (n + 2)) \right)} \nonumber\\
  &  & \cdot \left( n - n \cdot \exp (- \frac{2 \Omega}{T} (n + 2)) + (n + 2)
  \cdot \exp (- \frac{2 \Omega}{T} (n + 1)) - (n + 2) \cdot \exp (- \frac{2
  \Omega}{T}) \right)\nonumber\\
    \label{eq:dnu_system}
\end{eqnarray}
%\end{widetext}
with the derivatives of the asymptotes for $\nu=\mu$ being identically zero.
Differentiating by $T$ gives:
%\begin{widetext}
\begin{eqnarray}
  \frac{\partial \varepsilon}{\partial T} (k) & = & \frac{\varepsilon (k) -
  k^2 + \mu + \Omega}{T} + \left( a_2 \ast \frac{\partial \varepsilon /
  \partial T - \varepsilon / T}{1 + \exp ( \frac{\varepsilon}{T})} \right) (k)
  + \sum_{n = 1}^{\infty} \left( a_n \ast \frac{\partial \varepsilon_n /
  \partial T - \varepsilon_n / T}{1 + \exp ( \frac{\varepsilon_n}{T})} \right)
  (k)\nonumber\\
  &  & \nonumber\\
  \frac{\partial \varepsilon_n}{\partial T} (k) & = & \frac{\varepsilon_n
  (k)}{T} + \frac{1}{2 c} \left[ f \ast \left(  \frac{\partial
  \varepsilon_{n + 1} / \partial T - \varepsilon_{n + 1} / T}{1 + \exp (-
  \frac{\varepsilon_{n + 1}}{T})} + \frac{\partial \varepsilon_{n - 1} /
  \partial T - \varepsilon_{n - 1} / T}{1 + \exp (- \frac{\varepsilon_{n -
  1}}{T})} \right) \right] (k), \hspace*{1cm} (n \neq 1)\nonumber\\
  &  &\nonumber\\
  \frac{\partial \varepsilon_1}{\partial T} (k) & = & \frac{\varepsilon_1
  (k)}{T} + \frac{1}{2 c} \left[ f \ast \left(  \frac{\partial
  \varepsilon_2 / \partial T - \varepsilon_2 / T}{1 + \exp (-
  \frac{\varepsilon_2}{T})} - \frac{\partial \varepsilon_{} / \partial T -
  \varepsilon / T}{1 + \exp ( \frac{\varepsilon_{}}{T})} \right) \right] (k)\nonumber\\
  &  &\nonumber \\
    \lim_{n \rightarrow \infty} \frac{\partial \varepsilon_n / \partial T
  (k)}{n} & = &  0 \nonumber\\
  &  &\nonumber \\
  \frac{\partial \varepsilon_n^{\infty}}{\partial T} & = & \ln \left[ \left(
  \frac{1 - \exp (- \frac{2 \Omega}{T} (n + 1))}{1 - \exp (- \frac{2
  \Omega}{T})} \right)^2 - e^{- \frac{2 \Omega}{T} n} \right] +  \nonumber \\
  & &\frac{2 \Omega}{T} \frac{  2 \exp(- \frac{2 \Omega}{T}) \frac{\left( 1 - \exp (-
  \frac{2 \Omega}{T} (n + 1)) \right)^2}{\left( 1 - \exp (- \frac{2
  \Omega}{T}) \right)^3}- n \exp(- \frac{2 \Omega}{T}
  n) - 2 (n + 1) \exp(- \frac{2 \Omega}{T} (n +
  1))  \frac{1 - \exp (- \frac{2 \Omega}{T} (n + 1))}{\left( 1 - \exp (-
  \frac{2 \Omega}{T}) \right)^2} }{ \left( \frac{1 - \exp (- \frac{2
  \Omega}{T} (n + 1))}{1 - \exp (- \frac{2 \Omega}{T})} \right)^2 - e^{-
  \frac{2 \Omega}{T} n}} \nonumber\\
  \label{eq:dT_system}
\end{eqnarray}
%\end{widetext}
And by $c$:
%\begin{widetext}
\begin{eqnarray}
  \frac{\partial \varepsilon (k)}{\partial c} & = & - T \cdot \left(
  \frac{\partial a_2}{\partial c} \ast \ln \left[ 1 + \exp (-
  \frac{\varepsilon}{T}) \right] \right) (k) - T \sum_{n = 1}^{\infty} \left(
  \frac{\partial a_n}{\partial c} \ast \ln \left[ 1 + \exp (-
  \frac{\varepsilon_n}{T}) \right] \right) (k) \nonumber\\
  & + & \left( a_2 \ast \frac{\partial \varepsilon / \partial c}{1 + \exp (
  \frac{\varepsilon}{T})} \right) (k) + \sum_{n = 1}^{\infty} \left( a_n \ast
  \frac{\partial \varepsilon_n / \partial c}{1 + \exp (
  \frac{\varepsilon_n}{T})} \right) (k) \nonumber\ \\
  &  &  \nonumber\\
  \frac{\partial \varepsilon_n (k)}{\partial c} & = & \frac{T}{2 c} \left[
  \left( \frac{\partial f}{\partial c} - \frac{f}{c}
  \right) \ast \left( \ln \left[ 1 + \exp ( \frac{\varepsilon_{n + 1}}{T})
  \right] + \ln \left[ 1 + \exp ( \frac{\varepsilon_{n - 1}}{T}) \right]
  \right) \right] (k) \nonumber\\
  & + & \frac{1}{2 c} \left[ f \ast \left(  \frac{\partial
  \varepsilon_{n + 1} / \partial c}{1 + \exp (- \frac{\varepsilon_{n +
  1}}{T})} + \frac{\partial \varepsilon_{n - 1} / \partial c}{1 + \exp (-
  \frac{\varepsilon_{n - 1}}{T})} \right) \right] (k), (n \neq 1) \nonumber\ \\
  &  &  \nonumber\\
  \frac{\partial \varepsilon_1 (k)}{\partial c} & = & \frac{T}{2 c} \left[
  \left( \frac{\partial f}{\partial c} - \frac{f}{c}
  \right) \ast \left( \ln \left[ 1 + \exp ( \frac{\varepsilon_2}{T}) \right]
  + \ln \left[ 1 + \exp (- \frac{\varepsilon_{}}{T}) \right] \right) \right]
  (k) \nonumber\\
  & + & \frac{1}{2 c} \left[ f \ast \left(  \frac{\partial
  \varepsilon_2 / \partial c}{1 + \exp (- \frac{\varepsilon_2}{T})} -
  \frac{\partial \varepsilon_{} / \partial c}{1 + \exp (
  \frac{\varepsilon_{}}{T})} \right) \right] (k) \nonumber\ \\
  &  &  \nonumber\\
  \lim_{n \rightarrow \infty} \frac{\partial \varepsilon_n / \partial c
  (k)}{n} & = &  0 \nonumber\\
  &  &  \nonumber\\
 \frac{\partial \varepsilon^\infty_n }{ \partial c (k)} & = &  0 
  \label{eq:dc_system}
\end{eqnarray}

\end{widetext}

\section{Covariance under the parameter $c$}
\label{covariance}
The thermodynamics of the system depend on four parameters: $(c, \mu,
\Omega, T)$. Or we can easly deduce form (\ref{eq:system}) that they are
covariant under renormalisation by $c$, i.e.:
\begin{eqnarray}
  G (c, \mu, \Omega, T) & = & c^2 G (1, \frac{\mu}{c^2}, \frac{\Omega}{c^2},
  \frac{T}{c^2}) \nonumber\\
  n_i (c, \mu, \Omega, T) & = & c n_i (1, \frac{\mu}{c^2}, \frac{\Omega}{c^2},
  \frac{T}{c^2}), \hspace*{\fill} i = 1, 2 \hspace*{\fill} 
\end{eqnarray}
This allows one to reduce our parameter space to $\{\mu,
\Omega, T\}$ and put $c=1$ by default.

\end{appendix}

\bibliographystyle{apsrev}
\bibliography{2CBG}

\end{document}